# LES Study on the Mechanism of Vortex Rings behind Supersonic MVG with Turbulent Inflow

*Yonghua Yan[1], Caixia Chen[2], Ping Lu[3], Chaoqun Liu [4]*
University of Texas at Arlington, Arlington, Texas 76019, USA
cliu@uta.edu

In this study, we investigate the interaction between vortex rings behind MVG and the oblique shocks in the MVG controlled ramp flow at M=2.5 and $Re_\theta$=5760. Implicit large eddy simulation (ILES) method is used by solving the unfiltered form of the Navier-Stokes equations with the 5th order Bandwidth-optimized WENO scheme. The fully developed inflow is given by a series of turbulent profiles obtained from previous DNS simulation. It shows that the ring structure does not break down and keeps its topology after penetrating the strong shock wave and the oblique shocks is influenced a lot by the induced flow field from rings. The bump of the 3D shock wave surface is discovered and its mechanism is explained.

## Nomenclature

| | | |
|---|---|---|
| MVG | = | micro ramp vortex generator |
| $M$ | = | Mach number |
| $Re_\theta$ | = | Reynolds number based on momentum thickness |
| $h$ | = | micro ramp height |
| $\delta$ | = | incompressible boundary-layer nominal thickness |
| $x, y, z$ | = | spanwise, normal and streamwise coordinate axes |
| u,v,w | = | spanwise, normal and streamwise velocity |
| LES | = | large eddy simulation |
| $\Delta s_{ring}$ | = | averaged horizontal distance between rings |
| $V_{ring}$ | = | the horizontal component of the velocity of the vortex ring |
| $T_{ring}$ | = | averaged period of a ring passing $\Delta s_{ring}$ |
| $St_h$ | = | Strouhal number based on height of MVG $h/(U_\infty T_{ring})$ |

Subscript
| | | |
|---|---|---|
| $0$ | = | inlet |
| $w$ | = | wall |
| $\infty$ | = | free stream |

## I. Introduction

Shock-boundary layer interaction (SBLI) is a kind of problem which is frequently met in supersonic engine inlet flow and external flow. The interactions usually decrease the total pressure recovery, degenerate the shape factor of the supersonic boundary layer, and result in flow separation. Recently, to improve the circumstances, a sub-boundary layer control device called as micro vortex generator is developed as a new control device, which has the benefit of reduced drag and physical simplicity. It is considered to be a hopeful substitution for bleeding control. In contrast to the conventional counterpart,

---
[1] PhD Student, Department of Mathematics, Box 19408, Univ. of Texas at Arlington
[2] PhD Student, Department of Mathematics, Box 19408, Univ. of Texas at Arlington
[3] PhD Student, Department of Mathematics, Box 19408, Univ. of Texas at Arlington
[4] Professor, Department of Mathematics, Box 19408, Univ. of Texas at Arlington, AIAA Associate Fellow

the microramp vortex generator has the smaller size (approximately the height 20-40% of the boundary layer), longer streamwise distance for the vortices to remain in the boundary layer, and therefore better efficiency of the momentum exchange[9,10].

In Ref. 1 and 2, a series of computation were conducted on MVG controlled flow using LES at M=2.5 and $Re_\theta$=1440. Besides the traditional streamwise vortices model, a train of spanwise vortex rings generated continuously within the boundary of the momentum deficit was discovered. This new phenomenon is called "vortex rings". From the analysis of the mechanism for the vortex rings, we found that the existence of the high shear layer caused by the momentum deficit causes the corresponding Kelvin-Helmholtz instability, which further develops into a series of vortex rings1. Since the vortex rings can bring high energy to the lower boundary layer through "downwash" sweeps, corresponding mechanism should be considered at least as a part of the mechanisms of the flow control. Eight discoveries have been reported by Li and Liu[1,2]: 1.Spiral points and vortex structure around MVG; 2. Surface separation topology; 3. New model of five pairs of the vortex tubes around MVG (Figure 1); 4. Source of the momentum deficit; 5. Inflection surface, K-H instability and ring generation; 6. Ring-shock interaction and new mechanism of separation reduction; 7. 3-D structure of re-compression shock; 8.

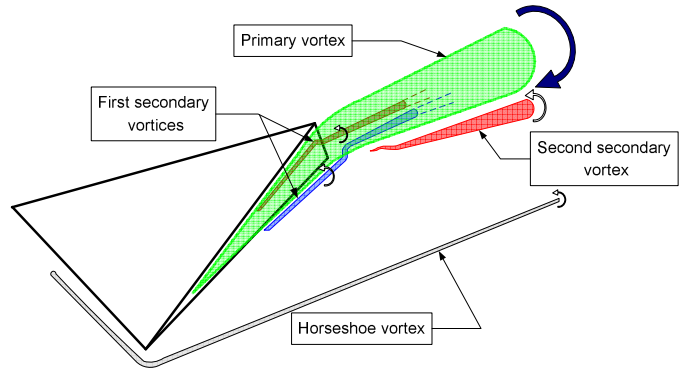

**Figure 1: The dynamic vortex model (Li and Liu[2])**

Influence of decline angles of the trailing edge of MVG.

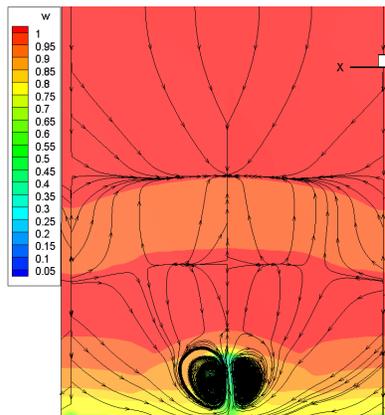

**Figure 2:    Structure of the deficit and the streamlines**

Vortex-shock wave interaction[3] has been studied a lot. Main concerned issues in this topic are: a) the deformation of the shock wave; b) the multistage features of the interaction caused by the vortex interaction with the primary shock and the reflection shocks, etc; c) the acoustic characteristics, which includes near- and far-field of acoustic, the dipole and quadruple acoustic pressure structure, etc.

Compared with the classic study Ref. 4-6, the vortex rings-shock interaction of the MVG controlled ramp flow is different case and could bring a new topic. The differences are: a) the interaction is a more complicated 3-D one than the 2-D counterpart, which happens between 3-D vortex rings and the oblique shock wave; b) the interaction happens within or close to the boundary layer and the separation region, where other flow structures exist like vortices with small scale besides the shock wave; c) the interaction is a continuous one, not a one-time event; d) besides the rings, components of the primary vortices still exist and make the interaction more complicated. Although differences are obvious, results obtained in the standard vortex ring-shock interaction can still give hints and suggestions to the current research.

## II. Numerical methods, grid generation and turbulent inflow

It is definitely needed to find physics of MVG for design engineers. RANS, DES, RANS/DES, RANS/LES, etc are good engineering tools, but may not be able to reveal the mechanism and get deep understanding of MVG. We need high order DNS/LES. A powerful tool is the integration of high order LES and experiment. In this paper, an approach named monotone integrated LES (MILES)[5,7] was adopted at Mach number 2.5 and $Re_\theta$=5760, in which the numerical dissipation is used as the sub-grid stress model.

Flows around MVGs are studied with back edge declining angle 70° (see Fig.3).The geometries for the cases are shown in Fig4 (in which $\delta_0$ represents the incompressible boundary layer nominal thickness). A general grid partition technique is used in this grid generation. According to experiments by Babinsky[9], the ratio h/$\delta_0$ of the models range from 0.3 to 1. The appropriate distance from the trailing-edge to the control area is around 19~56h or 8~19$\delta_0$. In this study, the height of MVG h is assumed to be $\delta_0$/2 and the horizontal distance from the apex of MVG to the ramp corner is set to be 19.5h or 9.75$\delta_0$. The distance from the end of the ramp to the apex is 32.2896h. The distance from the starting point of the domain to the apex of MVG is 17.7775h. The height of the domain is from 10h to 15h and the width of the half domain is 3.75h. As shown in Fig.4, three regions are divided as: the ramp region, MVG region and fore-region. Between each two regions, there is a grid transition buffer. Because of the symmetry of the grid distribution, only half of the grids need to be generated. The grid number for the whole system is: $n_{spanwise} \times n_{normal} \times n_{streamwise}$=128×192×1600. Using the inflow flow profile described in the next section, a data summary is given in table 1 about the geometric parameters of the grid system.

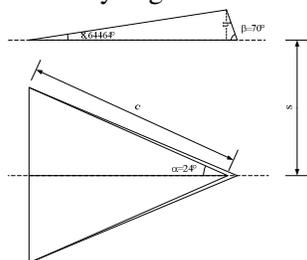 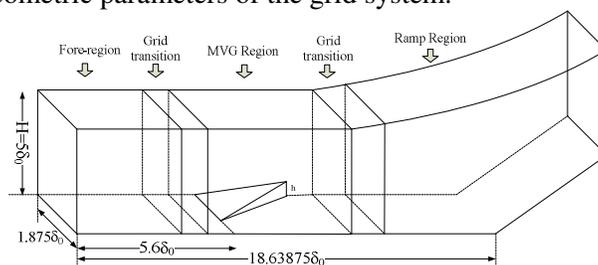

Figure 3: The sketch of MVG at $\beta$ =70°     Figure 4: The schematic of the half grid system

| $L_x$ | $L_y$ | $L_z$ | $\Delta x^+$ | $\Delta y^+$ | $\Delta z^+$ |
|---|---|---|---|---|---|
| 3.75$\delta_0$ | 5-7.5$\delta_0$ | 25.03355$\delta_0$ | 26.224 | 1.357-38.376 | 12.788 |

**Table 1. The geometric parameters for the computation**

The details about the geometric objects, grid generation, computational domain, etc, which are introduced in our previous paper, [2, 6] will not be repeated here.

The adiabatic, zero-gradient of pressure and non-slipping conditions are adopted at the wall. To avoid possible wave reflection, the non-reflecting boundary conditions are used on the upper boundary. The boundary conditions at the front and back boundary surfaces in the spanwise direction are treated as the periodic condition, which is under the consideration that the problem is about the flow around MVG arrays and only one MVG is simulated. The outflow boundary conditions are specified as a kind of characteristic-based condition, which can handle the outgoing flow without reflection [3].

New fully developed turbulent inflow boundary conditions are generated in this paper using following steps:
a) A turbulent mean profile is obtained from previous DNS simulation results[8] for the streamwise velocity (w-velocity) and the distribution is scaled using the local displacement thickness and free stream velocity. The basic transfer is based on the assumption that the same distribution exists between the relations of $U/U_e \sim y/\delta^*$. And the averaged streamwise velocity of MVG case can be reached by interpolation (3rd Spline interpolation); b) The pressure is uniform at inlet and has the same value as the free stream value. The temperature profile is obtained using Walz's equation for the adiabatic wall: first, the adiabatic wall temperature is determined using: $T_w = T_e(1 + r(\gamma-1)/2 \times M_e^2)$, where the subscript 'e' means the edge of the boundary layer and r is the recovery factor with value 0.9; next, the temperature profile is obtained by Walz's equation: $T/T_e = T_w/T_e - r(\gamma-1)/2 \times M_e^2 (U/U_e)^2$; c) The fluctuation components of the velocity are separated from the total velocity at every instantaneous data file (totally 20,000 files). And such fluctuations are rescaled in the same way. Because $\bar{T}/T_e = T_w/T_e - r(\gamma-1)/2 M_e^2 (\bar{U}/U_e)^2$, considering the non-dimensional form and ignore the $T_e$ and $U_e$, we get $d\bar{T} = -r(\gamma-1)M_e^2 U d\bar{U}$, or $\Delta T = -r(\gamma-1)M_e^2 U \Delta U$. Density fluctuation is determined by $\frac{\Delta \rho}{\bar{\rho}} = -\frac{\Delta T}{\bar{T}}$; d) Finally, the transformed parameters are

$u = U + \Delta u$, $v = V + \Delta v$, $w = \Delta w$, $\rho = \bar{\rho} + \Delta \rho$, $p = \frac{\rho T}{\gamma M^2}$, $T = \bar{T} + \Delta T$.

To check the flow properties before the MVG, we analyzed the relevant flow parameters on a spanwise cross section which is illustrated in Fig 5. The cross section is 11.97h ahead the apex of MVG. As a result, the displacement thickness $\delta^* = 0.371h$, the momentum thickness $\theta = 0.275h$, nominal boundary layer thickness $\delta = 2.36h$. Thus, we can obtain a shape factor H as about 1.35 which shows the flow before the MVG is fully developed turbulence flow.

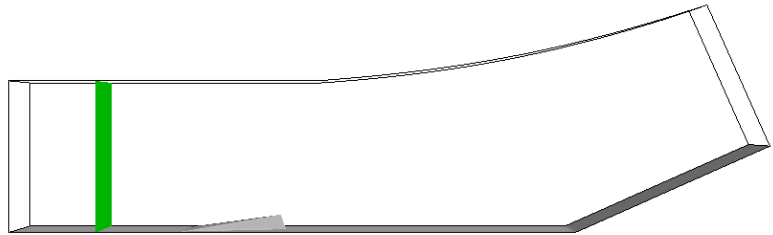

**Figure 5: The spanwise cross section on which the flow parameters are checked**

Fig. 6 shows the inflow boundary layer velocity profile in log-coordinates on the same cross section. There is a well-defined log region and the agreement with the analytical profile is well established. These results are typical for a naturally grown turbulent boundary layer in equilibrium.

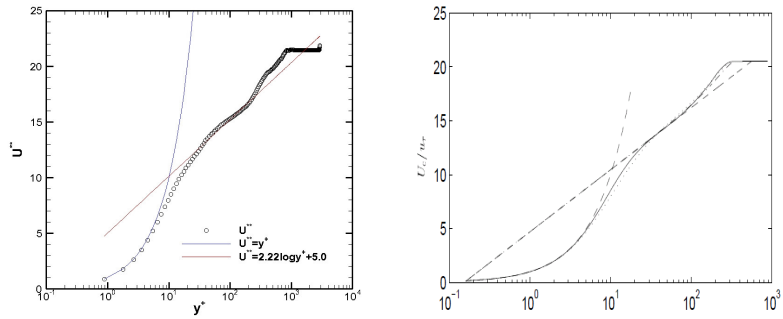

**Figure 6: Inflow boundary-layer profile comparison with Guarini et al's**

### III. Vortex ring structure

In Fig. 8 the iso-surface of $\lambda_2$ scalar field is given. It is very clear that there is a chain of vortex rings, which start from the trailing-edge of MVG. The rings are placed almost erectly and then are continuously distorted and enlarged while they are propagating downstream. These rings could dominant the mechanism of MVG for control of shock boundary layer interaction.

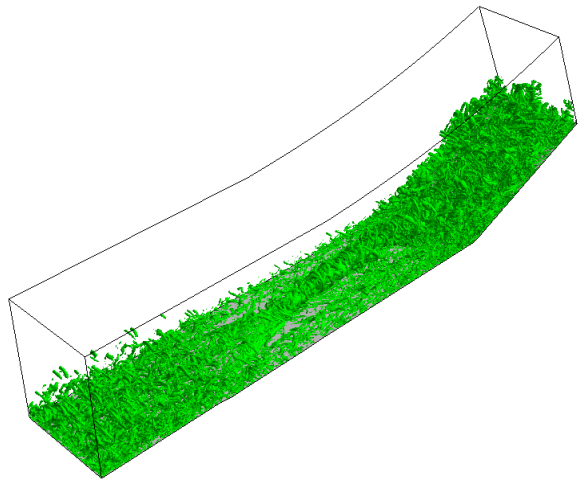

**(a) global view**

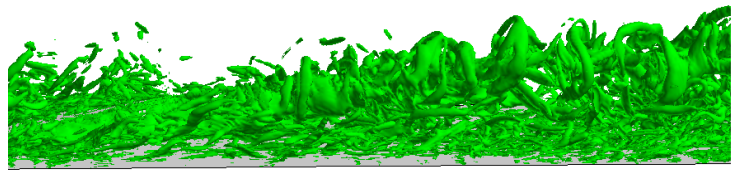

**(b) close-up view behind the MVG**
**Figure 8. Vortex rings shown by iso-surface of $\lambda_2$**

In Fig. 9, we demonstrate the instantaneous numerical schlieren picture at the central plane. From the figure, we can see many vortex rings appear in the circular shapes. After being informed the prediction of the vortex rings, the experimentalists in UT Arlington tried some techniques to validate the discovery.

They applied the techniques of the particle image velocimetry (PIV) and the acetone vapor screen visualization to track the movement of the flow. And more specifically the flash of a laser sheet is used to provide the light exposure at the time interval of micro seconds. In Fig.10, a typical image at the center plane is taken by using PIV and the acetone vapor (Lu et al [11]). It is clearly demonstrated that a chain of vortex rings exist in the flow field after the MVG! And these structures qualitatively resemble those in Fig. 9 by our LES.

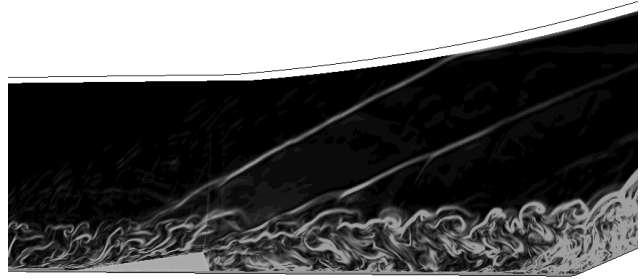

**Figure 9. The numerical shilieren at the center plane**

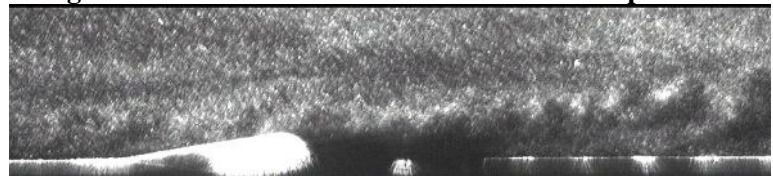

a) Using PIV

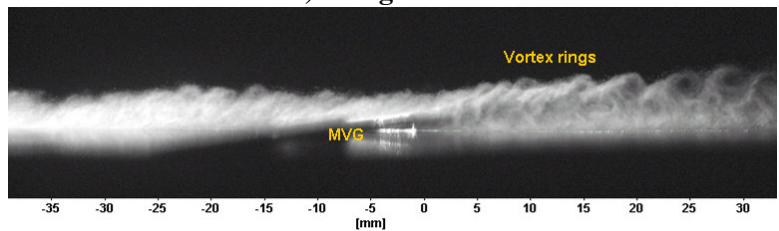

b) Using the acetone vapor

**Figure 10.    The laser-sheet flash image at the center plane (Lu et al 2010)**

Our numerical discoveries of the vortex ring structure are also confirmed by 3-D PIV experiment (Fig. 11) conducted by Sun et al at Delft University[12]. Compared with our numerical result in Fig. 12, we can find the similar distribution of streamwise ($\omega_z$) and spanwise vorticity ($\omega_x$) components which also proves the existence of ring structures. The Kevin-Helmhotz vortices part in Fig. 11 corresponds the ring head in Fig. 8. The underneath part which illustrated as streamwise vortices are two counter rotating primary vortices which are considered as the main source of the ring structure as explained later .

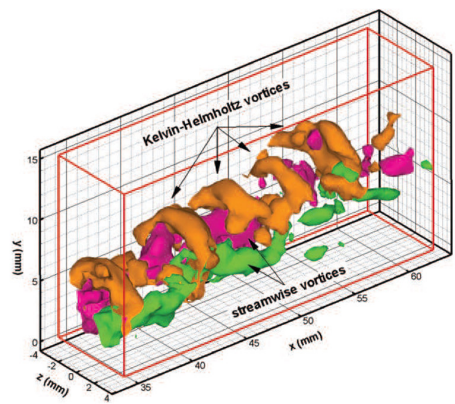

**Figure 11. K-H rings behind MVG by (Sun et al 2011)**

6
American Institute of Aeronautics and Astronautics

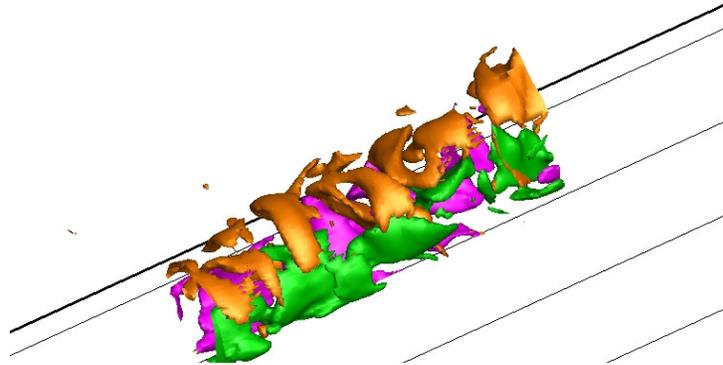

**Figure 12. Distribution of Kelvin-Helmholtz vortices and streamwise vortices from LES**

However the vorticity component which revolves towards the vertical direction($\omega_y$ in our case) is not shown in Fig. 11. If this missing part was provided, we can see the vortex ring structure clearly by the combination of all the components of vorticity as shown in Fig. 13 which is in accordance with the structure in Fig. 8.

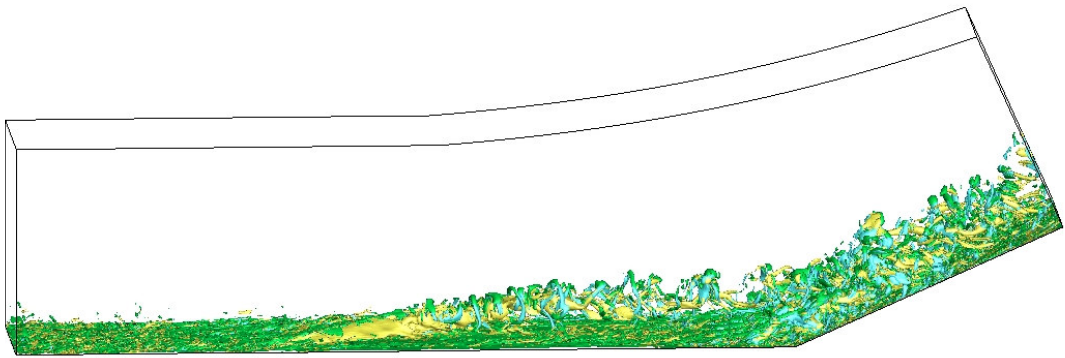

**(a) global view**

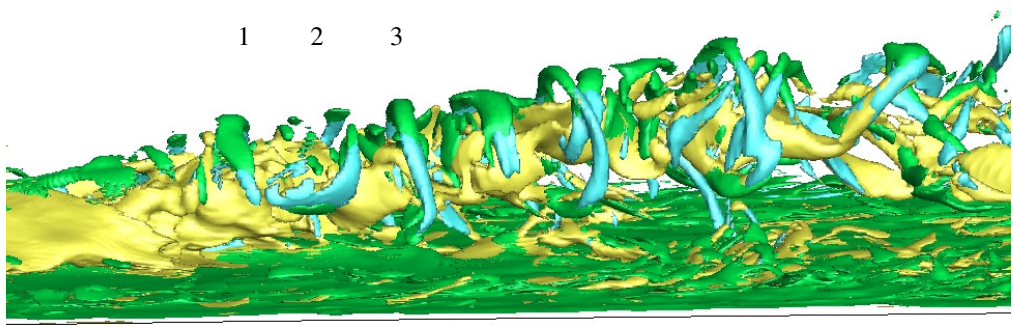

**(b) close-up view**
**Figure 13. Vortex rings shown by the components of vorticity**

### IV. Influence on ring structure by the interaction

In order to make further analyses, it is necessary to get the kinetic information of the vortex rings. According to the results of computation, it is found that vortex rings appear irregularly after the MVG and before the ramp. They are continuously distorted during propagating. So only Ring 3 of these

rings(marked in Fig. 13b) are checked since they are relatively regular at this stage. It shows that they almost propagate in the same speed and the averaged parameters are presented in table 1.

| $\Delta s_{ring}$ | $V_{ring,1}$ | $T_{ring}$ ($h/U_\infty$) | $St_h$ |
|---|---|---|---|
| *1.21h* | *0.78$U_\infty$* | *1.55* | *0.26* |

**Table 2. Characteristic parameters of vortex rings on the plate**

The shape of vortex rings is badly deformed before they penetrate the shock wave and travel along the ramp. But it is still possible to obtain the approximate speed of the two obvious vortex rings right in front of and behind the shock (see Fig. 14, marked as 1 and 2), the value of the streamwise velocity can be found as: Vring_1≈0.77$U_\infty$, Vring_2≈0.47$U_\infty$

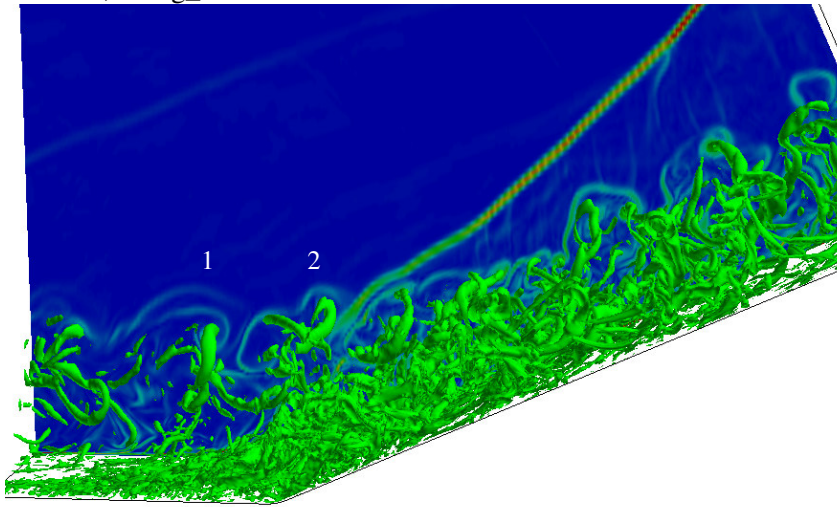

**Figure14.   The three vortex rings above the ramp for measurement**

For the second ring, since Vring_2 is only the component of the streamwise velocity, the real velocity will be about 0.68$U_\infty$ while it is evaluated along the direction of ramp. The value is consistent with the common knowledge, i.e., typical convective structures usually travel at a speed around 0.7$U_\infty$. For the first velocity Vring_1, the result has the same quantity level as well. We can also find that the total speed of the ring dose not change much when it penetrates the shock wave. A little decrease in the velocity may be caused by the interaction between ring and viscous sub-layer since the rings merge into the lower position in the boundary layer after they penetrate the shock. However, this phenomenon needs to be further verified.

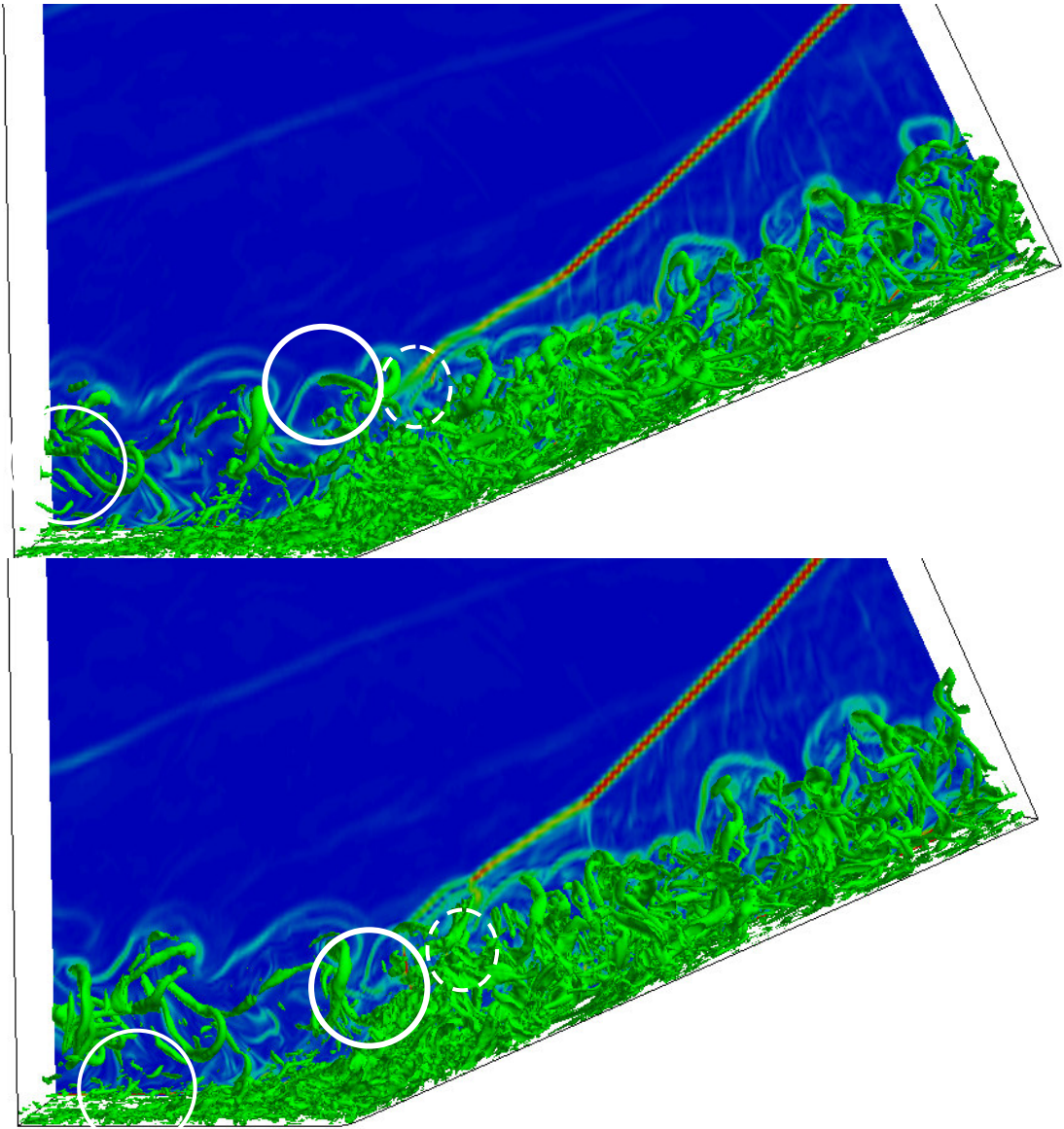

**Figure 15. The 3D view of the propagating rings at the ramp part at different time**

In Fig. 15, we tracked 3 different rings when they are moving. When the vortex ring is passing through the shock wave, the vortex structure is slightly distorted but not broken. The vortex ring does not break down by penetrating the strong shock wave and even keeps the topology very well. Since the vortex ring is distorted continuously after it is created (as we can find in Fig. 15, the first ring in circle), there's no conclusion that the interaction of vortex rings and shock makes the ring to be distorted. The distortion of rings could be induced by the particular flow field of inside and outside the ring structure, or the compressing effect of the ramp. In Fig. 15 and Fig. 16 we can also find that those vortex rings keep their existing shape after penetrating the shock wave and their topology are well maintained to the end of the ramp. That means in this 3D case, shock rarely affects the vortex structure.

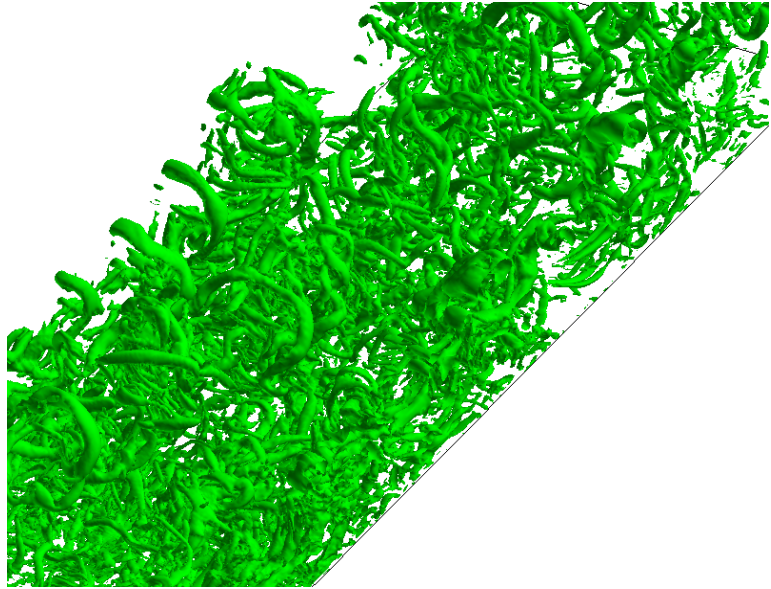

**Figure 16. The ring structure at the ramp part by λ2**

The interaction of vortex rings and shock wave can be explained with Richtmyer-Meshkov instability (RMI). It occurs when an interface between fluids of differing density is impulsively accelerated, e.g. by the passage of a shock wave. Since the vortex rings structure is a wave of differing densities, the process of passing the shock wave is the same with that in RMI theory. The key point of RMI is the baroclinic effection between $\nabla \rho$ and $\nabla p$. The quantity of $|\nabla \rho \times \nabla p|$ plays a very important role to the vorticity disturbance. Fig. 17 shows the iso-surface of $|\nabla \rho \times \nabla p|$ with different values, we can find that : a) The greater value of $|\nabla \rho \times \nabla p|$ only exists in a small area, thus the flow field, especially for the vortex structure, is rarely affected. b) The small value of $|\nabla \rho \times \nabla p|$ mainly happens in the bottom of the boundary layer, so the shock wave will not change the vortex rings a lot while those rings only exist on the upper side of the boundary layer. That may be the reason why the ring structure is so robust and never break down even when it penetrates the strong shock wave.

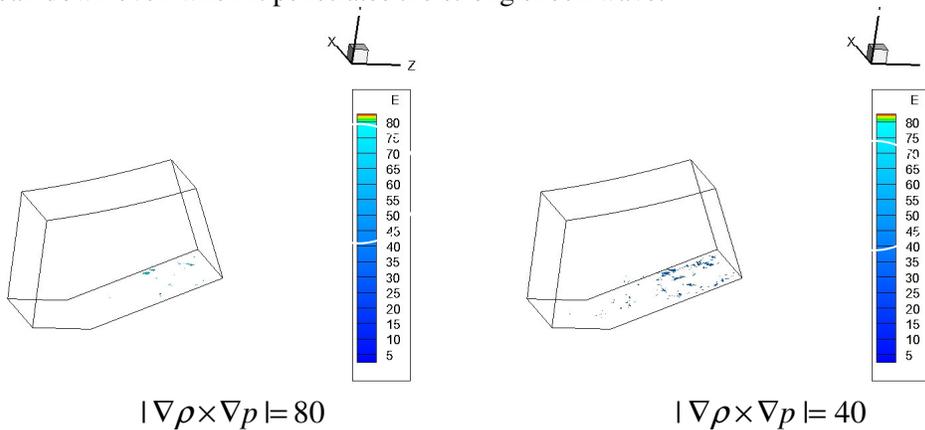

$|\nabla \rho \times \nabla p| = 80$  $\qquad\qquad$  $|\nabla \rho \times \nabla p| = 40$

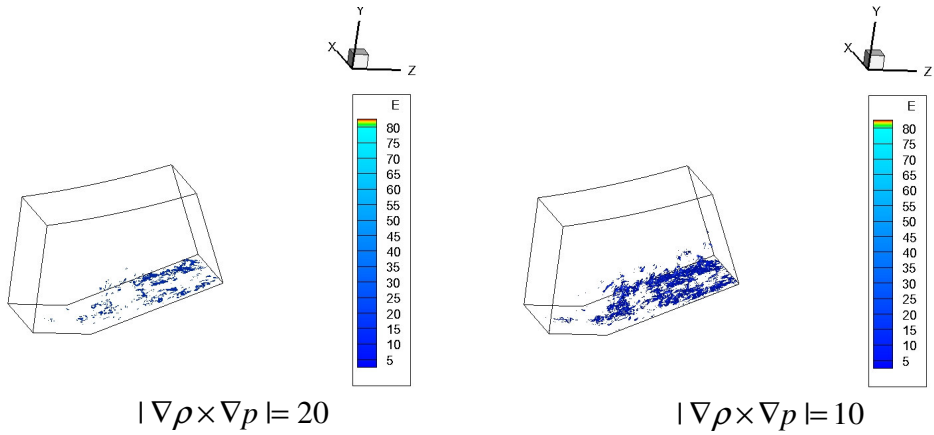

$|\nabla\rho \times \nabla p|= 20$          $|\nabla\rho \times \nabla p|= 10$

**Figure 17.** Iso-surface of $|\nabla\rho \times \nabla p|$ in the ramp part

## V. Influence on oblique shock wave

As newly found structure, the string of vortex rings are fully 3D structures. Thus the interaction with shock wave have different features from classic 2-D ones.

To better illustrate the process, two kinds of sections are investigated: the central plane and then a series of spanwise computational planes with constant streamwise index. In Fig. 18, The gradient fields of density $|\nabla\rho|$ and pressure $|\nabla p|$ at two different moments are employed to analyze the interaction. Fig. 19 gives the contour of $|\nabla\rho|$ and $|\nabla p|$ of spanwise computational planes at the same three successive moments. In order to get the overall understanding of the interaction, it is also necessary to directly give the 3-D shape of the shock wave. In Fig. 20, six snapshots are obtained using the iso-value of pressure. The value of pressure is selected to make the inner shock layer be exposed and seen from outside. The shape of the strong shock wave is given by $|\nabla p|$ in Fig 21.

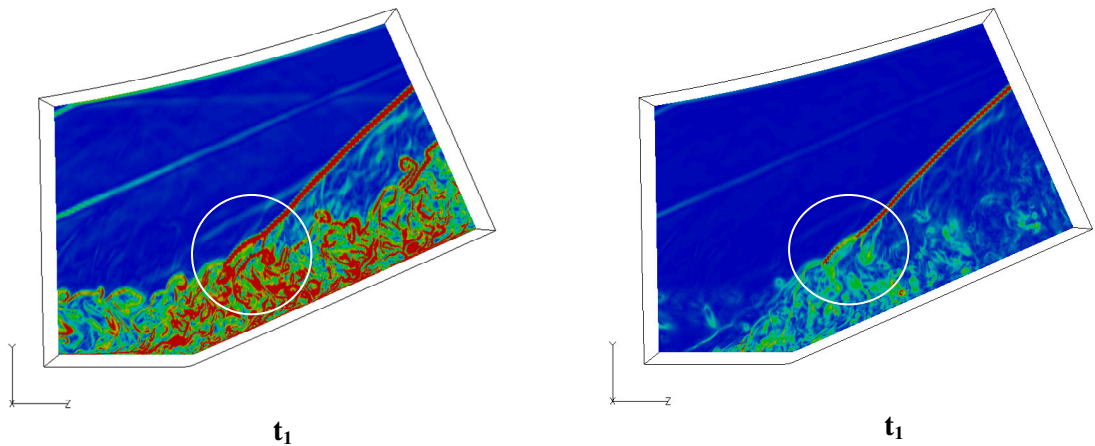

$t_1$          $t_1$

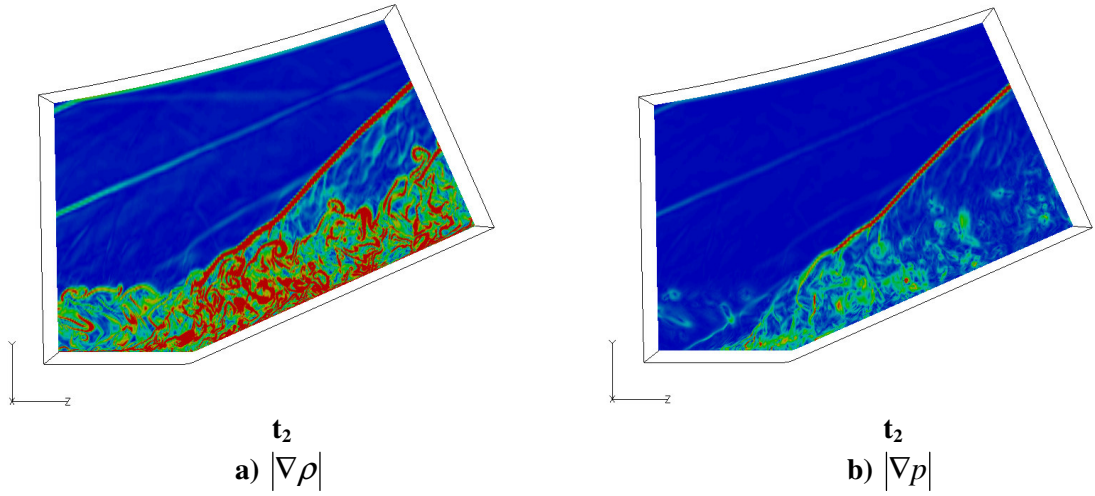

**t₂**
**a)** $|\nabla\rho|$

**t₂**
**b)** $|\nabla p|$

**Figure 18.** Contours of $|\nabla\rho|$ and $|\nabla p|$ at two moments at the center planes

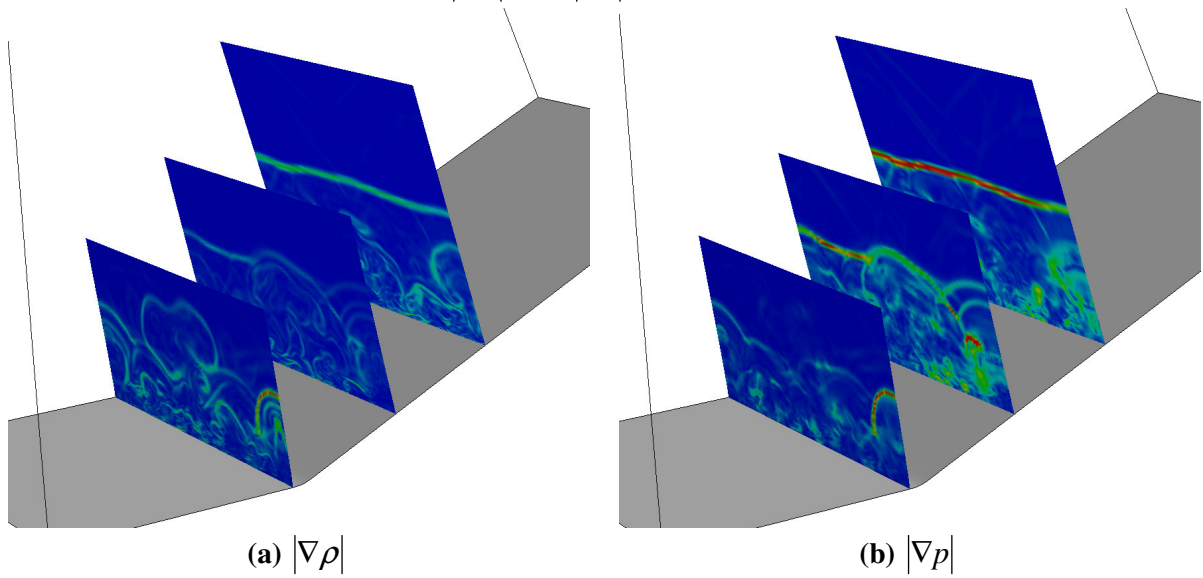

**(a)** $|\nabla\rho|$

**(b)** $|\nabla p|$

**Figure 19.** The contours of $|\nabla\rho|$ and $|\nabla p|$ at spanwise sections and at four successive moments

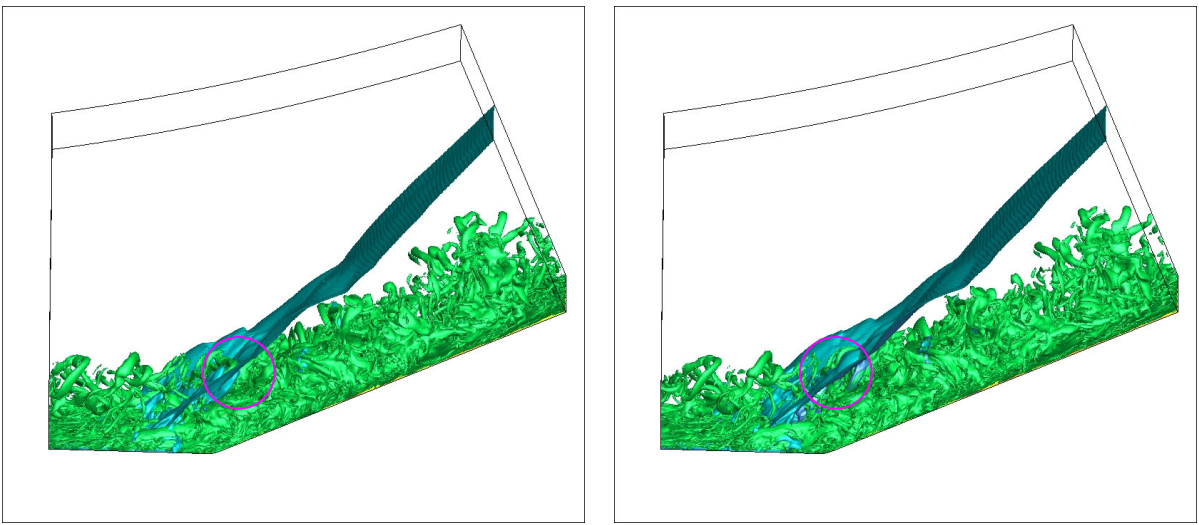

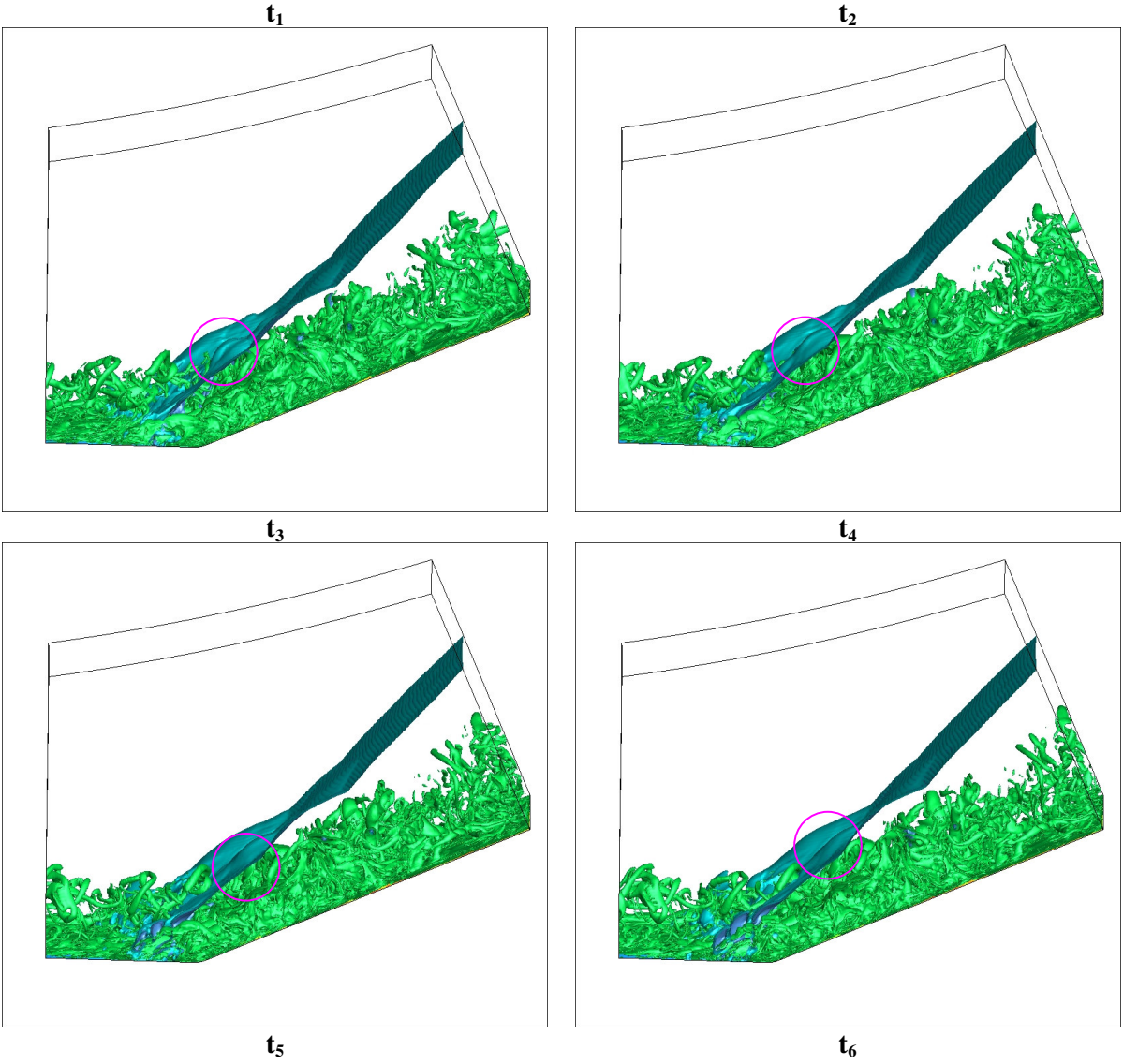

**Figure 20.   The shock wave shape by the iso-surface of pressure and ring structure by $\lambda_2$**

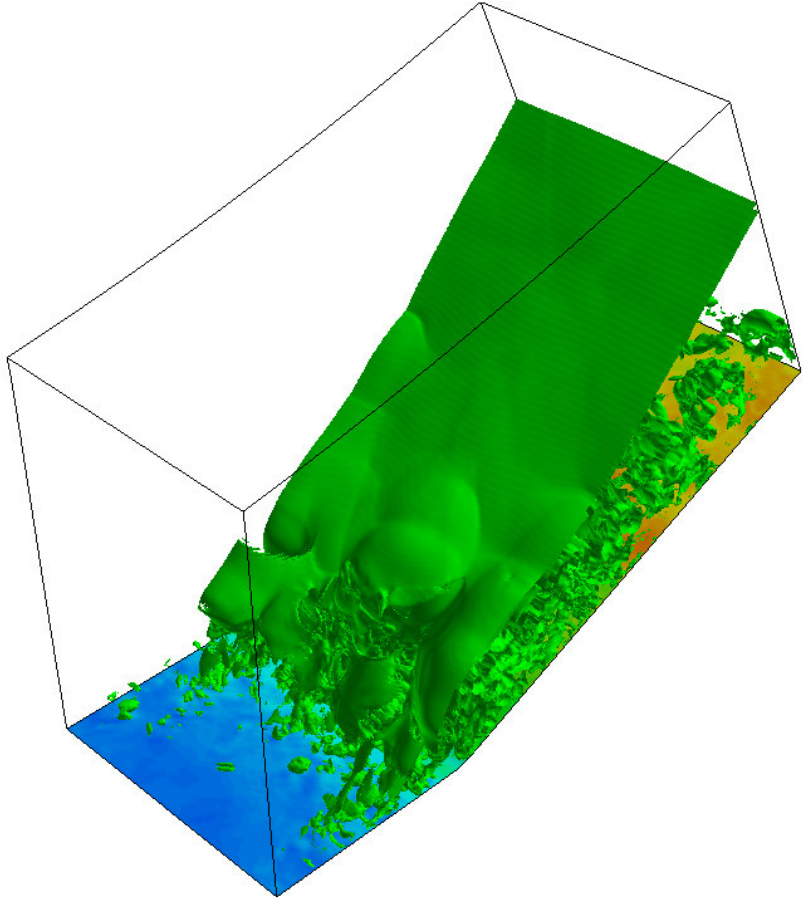

**Figure 21.   The 3D shock wave shape by $|\nabla p|$**

From these figures, the following features can be observed:

1) In Fig. 18, it demonstrates that there are at two layers of shock wave or wave structures after the corner: upper one is the original but quite weaker separation/reflection shock; the other is the stronger interacting shock wave caused by the vortex rings. These two layers of shocks will merge into one shock wave afterwards, i.e., the oblique shock caused by ramp.

2) Comparing Fig. 18(a) and (b), it can be found that there is a slip line under the curved shock in the contour of $|\nabla \rho|$, which cannot be found in the contour of $|\nabla p|$. The slip line indicates the density change across the line, while the pressure keeps the same at the both sides of the line. After checking the movie about $|\nabla \rho|$, the slip line comes from the original connections between vortex rings.

3) In Fig. 19, we can found that when a vortex ring penetrates into the shock wave, the interaction part of the shock is distorted and its intensity is also reduced a lot.

4) In Fig. 20 and Fig. 21, it shows that the interaction between shock wave and vortex rings is fully 3D. The flow field lost its symmetry and it can be clearly illustrated by the shape of shock wave. Multilayer structures and the bump like 3-D shape is one of the typical characteristics of the shock-vortex interaction.

5) In Fig. 21, it can be observed that, at the corner of the ramp, there is no obvious sign of shock waves, and the original shock wave retreats to a downstream position on the ramp. This result shows that the separation/reflection shock wave is eliminated near the corner. We can see that it is induced by the interaction with vortex rings and the viscous sub-layer (Fig. 20). This kind of phenomenon is a complicated turbulence-shock wave interaction which should be studied further.

6) When the vortex rings pass through the shock wave (Fig. 20 at time $t_1$), the shock wave will be distorted like a bump (Fig. 20 at time $t_3$), and the bump will be gradually smoothed when propagating downstream. We can clearly see that this shape is caused by the vortex ring-shock wave interaction. When the vortex enters the shock wave and moves away from it, the distortion subsides quickly. The subsequent incoming vortices will repeat the process. At the fifth section in Fig. 20, the bump shape is less observable, which indicates the vortex combination moves far below the shock wave and have less influence on the separation/reflection shock wave.

To explain the mechanism of the bump shape of the 3D shock wave, in Fig. 22, the second ring in Fig. 14 (marked as 2) is taken out and the flow filed on a spanwise cross section, which pass through the center of this ring, is plotted. The velocity field is the tangent projection of the 3D velocity vector distribution on the cross section which ignores the velocity component along the ramp direction. The projection of free stream velocity moves towards the ramp since the cross section is vertical to the ramp surface. Meanwhile, as noticed in Fig. 15, when the vortex rings enter the shock wave, they incline to some degree to the ramp. What is very interesting is that the vector field, shown in the section, clearly demonstrates the upward induction of the flow at the center position of the ring. The inducted flow will interact with the incoming free stream and make the surface of the interaction which is the location of the shock wave to be an obvious arc-like boundary.

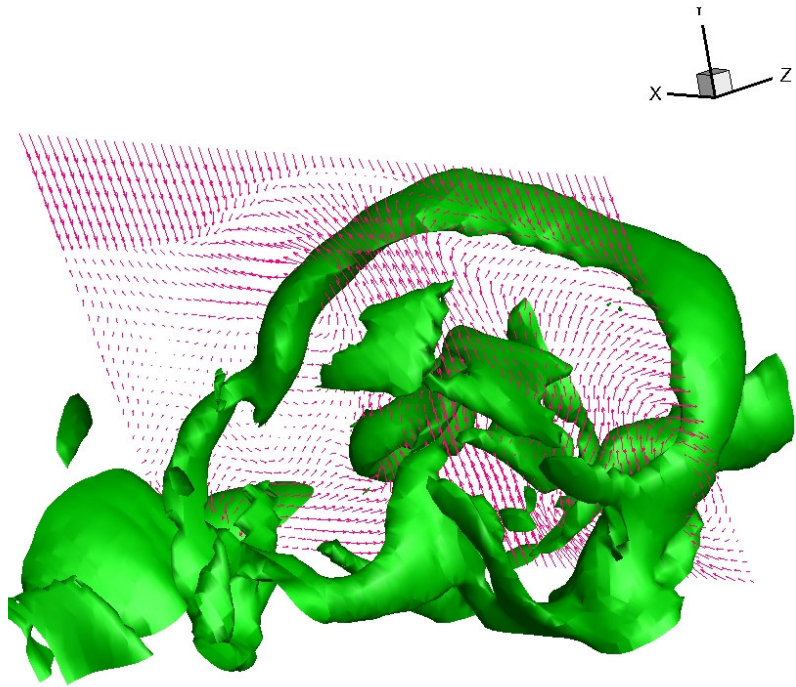

**Figure 22. The vector field at a spanwise section with vortex structure shown by total vorticity**

However, the interaction between the vortex ring structure and shock wave is a new interesting phenomenon. The existence of the interaction needs the verification of experiment, and the mechanism is still under investigation.

## VI. Conclusion

The interaction between vortex ring structure and the oblique shock by the MVG controlled ramp flow at M=2.5 and Re=5760 is studied in this paper. It shows that the ring structure does not break down and keeps its topology after penetrating the strong shock wave, while the strong oblique shock is influenced a lot by the induced flow field from vortex rings. The mechanism during the interaction is also discussed.

## Acknowledgements

This work was supported by The Department of Mathematics at University of Texas at Arlington . The authors are grateful to Texas Advanced Computing Center (TACC) for providing computation hours. This work is accomplished by using Code DNSUTA which was released by Dr. Chaoqun Liu at University of Texas at Arlington in 2009.